**Weak antilocalization beyond the fully diffusive regime in Pb$_{1-x}$Sn$_x$Se topological quantum wells**

Jiashu Wang,[1] X. Liu,[1] C. Bunker,[1] L. Riney,[1] B. Qing,[1,2] S.K. Bac,[1] M. Zhukovskyi,[3] T. Orlova,[4] S. Rouvimov,[3,4] M. Dobrowolska,[1] J. Furdyna,[1] B.A. Assaf[1]

[1] *Department of physics, University of Notre Dame, Notre Dame IN 46556*

[2] *School of Science, Xi'an Jiaotong University, Xi'an Shaanxi, 710049, China*

[3] *Department of Electrical Engineering, University of Notre Dame, Notre Dame IN, 46556*

[4] *Notre Dame Integrated Imaging Facility, University of Notre Dame IN, 46556*

**Abstract.** We report the measurements and analysis of weak antilocalization (WAL) in Pb$_{1-x}$Sn$_x$Se topological quantum wells in a new regime where the elastic scattering length is larger than the magnetic length. We achieve this regime through the development of high-quality epitaxy and doping of topological crystalline insulator (TCI) quantum wells. We obtain elastic scattering lengths that exceeds 100nm and become comparable to the magnetic length. In this transport regime, the Hikami-Larkin-Nagaoka model is no longer valid. We employ the model of Wittmann and Schmid to extract the coherence time from the magnetoresistance. We find that despite our improved transport characteristics, the coherence time may be limited by scattering channels that are not strongly carrier dependent, such as electron-phonon or defect scattering.

## I. Introduction

The Z$_2$ topological insulator class – discovered more than a decade ago – has delivered several exciting fundamental advances and promises applications in quantum computing and spintronics. Recently, a crystalline symmetry protected topological phase has been identified in the IV-VI Pb$_{1-x}$Sn$_x$Te and Pb$_{1-x}$Sn$_x$Se material class. The two systems have been shown to host four Dirac cones per surface.[1,2,3] Crystalline symmetry and valley degeneracy in these materials were proposed to yield a quantum Hall ferroelectric and a quantum anomalous Hall effect with a high-Chern number.[4,5,6] These research frontiers have remained unexplored. Their realization requires IV-VI topological quantum wells (QWs) with low carrier density and a controlled interface chemistry and band alignment (shown in Fig. 1(a,b)). Such single QWs are not readily available.

Most previous studies reporting transport measurements on TCIs have studied weak antilocalization (WAL) in SnTe, a material known to host carrier densities on the order of 10$^{20}$cm$^2$/Vs and mobilities on the order to 10-100cm$^2$/Vs.[7,8,9,10] Trivial PbSe, PbTe and the ternary non-trivial Pb$_{1-x}$Sn$_x$Se and Pb$_{1-x}$Sn$_x$Te have been synthesized with superior quality in the form of bulk epilayers and heterostructures.[11,12,13,14,15,16,17] However, there has not been any effort dedicated to the synthesis of high-quality single QWs of the non-trivial class of those materials. Thus, while WAL has been extensively studied in TIs and spin-orbit coupled systems, it remains underexplored in TCIs – especially in high mobility samples. WAL is a very useful tool to probe the properties of Dirac electrons,[18,19,20,21] particularly the Berry phase and how it is impacted by different symmetry breaking perturbations.

This work reports the realization of single QWs of TCI Pb$_{0.7}$Sn$_{0.3}$Se with carrier density lower than 10$^{13}$cm$^{-2}$ and mobility exceeding 10000 cm$^2$/Vs. This improved quality allows us to reach a new quantum coherent regime where the elastic scattering length $L_e$ exceeds the magnetic length ($L_B^2 = \hbar/2eB$) for $B < 0.1T$. In this limit, the well-known Hikami-Larkin-Nagaoka (HLN) model[22] is no longer valid. We thus implement a quantum coherent transport model developed by Wittman and Schmid (WS)[23] that remains valid beyond the fully diffusive regime to extract the inelastic scattering time $\tau_\phi$ and the Thouless coherence length. We show that the HLN model tends to underestimate degeneracies and overestimates the Thouless length $L_{Th}$ when compared to the WS model, but still reliably captures its decay versus temperature. Importantly, both models indicate that while $L_{Th}$ exceeds $1\mu m$ in low density QWs, its decay is also enhanced. Our results suggest that scattering channels that are not strongly carrier density dependent dominate electron decoherence in the IV-VI system. Most notably, electron-phonon or long-range electron-defect scattering play dominant role.

## II. Experimental Results
### A. Growth and characterization

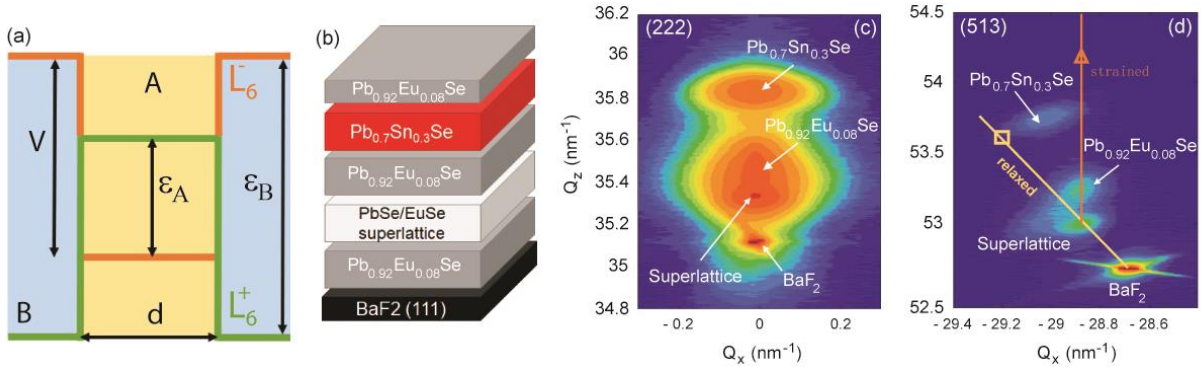

**FIG 1**. Basic properties for Pb$_{1-x}$Sn$_x$Se (111) Quantum Well. (a) Energy diagram of the topological QW, showing the symmetric confinement potential and the inversion of the bulk conduction and valence bands in the well. $\varepsilon_{A/B}$ denote the well (A) and barrier (B) energy gap respectively. $V$ denotes the band offset at the interface and $d$ is the thickness of the well. A band inversion occurs at the interfaces. (b) Structure diagram of quantum well. (c) X-ray diffraction reciprocal space map of (222) plane showing peaks attributed to the well, barrier, superlattice and substrate (d) RSM of (513) plane. It is clear that the superlattice is completely relaxed with respect to BaF$_2$ but it yields a slight in-plane tensile strain to Pb$_{0.7}$Sn$_{0.3}$Se. The expected fully relaxed and fully strained line of Pb$_{0.7}$Sn$_{0.3}$Se are marked by (□, Δ) respectively.

Quantum wells of Pb$_{1-x}$Sn$_x$Se with x≈0.3±0.03 oriented in the (111) direction are grown by molecular beam epitaxy (MBE) on (111)-BaF$_2$ substrates. A buffer layer structure is utilized to reduce the lattice mismatch between the well (a=6.085Å) and the BaF$_2$ substrate (a=6.196Å). The buffer layer consists of a starting buffer layer of Pb$_{0.92}$Eu$_{0.08}$Se (50nm, a=6.130 Å) followed by a 50 short-period superlattice of PbSe (1.6nm)/EuSe(1nm). The purpose of this initial structure is to suppress the propagation of lattice dislocations during growth and to maintain a smooth surface prior to the growth of the QW. A Pb$_{0.92}$Eu$_{0.08}$Se barrier (45nm) is then grown on the superlattice followed by the Pb$_{0.7}$Sn$_{0.3}$Se well (55±5nm) and a final capping layer Pb$_{0.92}$Eu$_{0.08}$Se (35nm) is synthesized on top. A schematic of the structure is shown in Fig. 1(b). High resolution X-ray diffraction (HRXRD) and transmission electron microscopy (Appendix A) are performed to characterize the heterostructures.

HRXRD is performed around the symmetric (222) node and the asymmetric (513) node. The resulting reciprocal space maps are shown in Fig. 1(c,d) respectively. Along the growth direction (q$_z$ || (111)), the Bragg peak resulting from the well is seen at the highest Bragg angle, clearly separated from the superlattice and buffer layer peaks. The Pb$_{0.7}$n$_{0.3}$Se peak appears at an angle slightly higher than what is expected from the bulk lattice constant, suggesting that the layer maintains a slight residual lattice strain that is tensile in the plane but compressive out-of-plane. The (513) space map shown in Fig. 1(d) confirms that while the buffer heterostructure is completely relaxed with respect to the BaF$_2$ substrate, the Pb$_{1-x}$Sn$_x$Se well is only partially relaxed with respect to the Pb$_{0.9}$Eu$_{0.08}$Se buffer and the superlattice. Bi doping is used to tune Fermi energy of the Pb$_{1-x}$Sn$_x$Se well, using an approach similar to what is already achieved in previous work.[24,25] A series of four samples with varying Bi content are grown. Their characteristics are summarized in table 1.

| Sample | Bi cell | $\rho$ [μΩ.cm] | Density [cm$^{-2}$] | Mobility[cm$^2$/Vs] | $L_e = v_F \tau_e$ | $B_{tr} = \dfrac{\hbar}{2eL_e^2}$ |
|---|---|---|---|---|---|---|
| QW-1 | No Bi | 2.3 | 3.6×10$^{13}$ | 4130 | 80 nm | 530 Oe |
| QW-2 | 305°C | 3.8 | 3.1×10$^{13}$ | 2950 | (52±4) nm | (1220±180) Oe |
| QW-3 | 310°C | 4.6 | - 6.6×10$^{12}$ | 11300 | 133 nm | 185 Oe |
| QW-4 | 320°C | 3.1 | - 1.3×10$^{13}$ | 8630 | (114±8) nm | (252±35) Oe |

**Table 1.** Sample information. From QW-1 to QW-4 we gradually increase the Bi doping density. The carrier density, mobility, mean free path and characteristic transport field are shown. They are all measured at 4.2K and found to not vary much between 4.2K and 10K.

### B. Electrical transport measurements

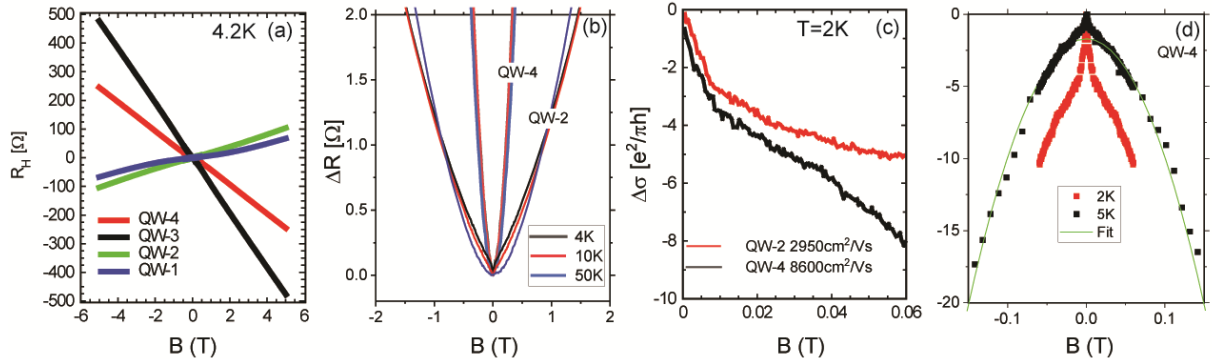

**FIG 2.** (a) Hall resistance of four QW with different Bi doping density. (b) Change in resistance versus magnetic field measured in QW-2 and QW-4 at different temperatures. (c) Magnetoconductance at low field showing weak antilocalization effect in QW-2 and QW-4 at 2K. (d) Magnetoconductance for QW-4 showing a coexisting cusp and parabolic field dependence. A curve fit is shown in green and yields a parabolic contribution equal to $820 \pm 30 e^2/(\pi h T^2) B^2$.

Electrical Hall effect measurements performed on this sample series up to 5T and down to at 4.2K are shown in Fig. 2(a). A control sample of Pb$_{0.9}$Eu$_{0.08}$Se/(PbSe,EuSe)/Pb$_{0.9}$Eu$_{0.08}$Se is checked to confirm that the buffer heterostructure is electrically insulating. Its resistance is found to exceed 33MΩ at room temperature, four orders of magnitude higher than with the QW. Pristine Pb$_{0.7}$Sn$_{0.3}$Se is p-type with a Hall concentration close to 3.6×10$^{13}$ cm$^{-2}$ likely due to group IV atom vacancies or substitutional defects. Bi is introduced into the lattice by co-evaporation during synthesis of the quantum well. Bi is a known donor in IV-VI materials and can alleviate defect-induced p-doping. The lowest 2D density that we reach is 6×10$^{12}$ electrons/cm² with a mobility that is above 11000 cm²/Vs. Table 1 summarizes the carrier density and mobility ($\mu$) extracted for four samples along with the corresponding elastic scattering length $L_e$ and characteristic transport field $B_{tr}$. The computation of $L_e$ requires a knowledge of the effective Fermi velocity $v_f$ and carrier effective mass $m$. Since the energy bands in IV-VI quantum wells are highly non-parabolic,[26,17] both quantities depend on the Fermi energy. In Appendix B and C, we discuss our determination of those two quantities given that multiple QW subbands are occupied.[27,28] After $v_f$ and $m$ are found, we have:

$$L_e = v_f \tau_e = \frac{v_f \, \mu \, m}{e}$$

$\tau_e$ is the elastic scattering time. The magnetoresistance (MR) of two samples (QW-2, QW-4) that this work will center on is shown in Fig. 2(b). The MR is dominated by a non-parabolic behavior at fields of order 1T; it is parabolic at intermediate fields and exhibits a clear cusp at low magnetic field (Fig. 2(c)). The high field MR is stronger in the higher mobility QW4 but can be consistently decoupled from the field dependence of the cusp. This is shown in Fig. 2(d) where a curve fit allows us to determine the parabolic part of the magnetoconductivity at intermediate fields for which WAL is nearly saturated. In both samples, the low field cusp characteristic of weak antilocalization (WAL) is observed at low temperatures, as typically seen in materials with strong spin-orbit coupling[29,30] and topological materials.[31,32,33] The cusp yields a magnetoconductance on the order of e²/h while the zero field Drude conductance is on the order of 100-1000e²/h (in QW4 $\sigma_{2D}(0) = 330 e^2/h$ at 2K) further confirming that it is due to the correction from quantum interference of backscattered electrons.[34] The field dependence of the resistance due to WAL yields the coherence length (or time) of charge carriers in the system.

In TIs, the WAL effect is particularly interesting to measure, as it can be a direct probe of the spin-momentum locking and the Berry phase of topological surface states, when the Fermi level is not too far from the Dirac point. In conventional semiconductors with strong spin-orbit coupling, WAL is generally accompanied by a crossover to weak localization (WL) as a function of magnetic field determined by the strength of the spin-orbit relaxation length. In TIs, WL is not allowed, as long as the Dirac surface states are not gapped, and spin-momentum locking is preserved thereby suppressing all backscattering. In TCI QWs, valley degeneracy can additionally yield a significant enhancement of the WAL correction.

### C. Modeling weak antilocalization

The well-known Hikami-Larkin-Nagaoka model[22,35](HLN) that describes the interference of electrons in the diffusive limit is used to extract the Thouless coherence length. At low magnetic fields, the quantum correction to the conductivity in this model in the limit of strong spin-orbit coupling applies to topological insulators and is given by:

$$\Delta \sigma_{2D} = \frac{\alpha e^2}{\pi h} \left[ \psi \left( \frac{\hbar}{4eBL_{Th}^2} + \frac{1}{2} \right) - \ln \left( \frac{\hbar}{4eBL_{Th}^2} \right) \right] + \beta B^2 \quad Eq.\,(1)$$

$\alpha = -1/2$ per WAL channel, a channel defined as a contribution from a separate valley or subband. $\psi$ is the digamma function. $L_{Th} = \sqrt{D\tau_\phi}$ is the Thouless coherence length related time and $D$ is the diffusion constant. $\beta B^2$ is added to account for cyclotronic magnetoresistance. $\beta$ is determined in Fig. 2(d) independently to best fit the parabolic part of the magnetoconductivity at intermediate field.

We note that the HLN model is valid only if $L_{Th} \ll L_e$ and if the applied field $B \ll B_{tr}$. For QW with high mobility such as QW-4, we find that $B_{tr} = 252\ Oe$, of the same order of magnitude as the applied field. When $B \sim B_{tr}$, the number of collision events per trajectory approaches 1. We are likely at the limit of validity of the HLN model.[22,36]

In conjunction with this model, we utilize a model developed by Wittmann and Schmid[23] (WS) that remains valid beyond the diffusive regime, to independently extract the coherence time $\tau_\phi$. The WS model[23] is implemented by computing the quantum correction to conductivity from interfering electron wavepackets. In general, the quantum coherent correction to the conductivity is given by:[23]

$$\Delta\sigma = -\frac{e^2}{\pi h} F(b,\gamma) \quad Eq.(2)$$

$$\gamma = \frac{\tau_e}{\tau_\phi} = \frac{B_\phi}{B_{tr}} \quad \text{where } B_\phi = \frac{\hbar}{4eD\tau_\phi}$$

$$b = \frac{B}{B_{tr}(1+\gamma)^2}$$

$F(b,\gamma)$ (<0) is related to the backscattering probability and given by

$$F(b,\gamma) = b\sum_{n=0}^{\infty} \frac{\varphi_n^3}{1-\varphi_n} - \ln(\frac{1+\gamma}{\gamma})$$

$$\varphi_n(b) = \int_0^\infty dt\, e^{-t-\frac{bt^2}{4}} L_n\left(\frac{bt^2}{2}\right)$$

$\varphi_n$ is the probability of a particle going back to the origin after $(n-1)$ scattering events. $L_n$ is the n-th Laguerre polynomial. $\Delta\sigma$ in Eq. (2) is thus a positive magnetoconductance resulting from weak localization (WL). The WS model is only valid if spin effects are neglected. Zduniak, Dyakonov and Knap[37] have considered the role of spin orbit coupling in modifying the quantum correction given by WS. The spin-orbit coupling is treated as a spin-relaxation in their model. As a result, they get

$$\Delta\sigma(B) = -\frac{e^2}{\pi h}\left[F(b,\gamma_1) + \frac{1}{2}F(b,\gamma_2) - \frac{1}{2}F(b,\gamma)\right] \quad Eq.(3)$$

Where, spin relaxation is introduced via the relaxation time $\tau_{SO}$ and the corresponding field $B_{SO}$ so that:

$$\gamma_1 = \frac{B_\phi + B_{so}}{B_{tr}}, \quad \gamma_2 = \frac{B_\phi + 2B_{so}}{B_{tr}}$$

In the presence of spin-momentum locking, in the topological crystalline insulator, we assume fast spin relaxation so that $B_{so}$ is very large at least of the same order as $B_{tr}$. This is equivalent to the simplectic limit of the HLN model, known to hold in the case of topological insulators[18]. In this limit, since $B_\phi \ll B_{tr}$ and $B_\phi \ll B_{so}$, $\gamma_1 \sim \gamma_2 \sim 1$ and we can neglect $F(b,\gamma_1)$ and $F(b,\gamma_2)$ (see appendix D). Hence, In the limit of strong spin-orbit coupling or spin-momentum locking, Eq. (3) reduces to:

$$\Delta\sigma(B) = +\frac{e^2}{2\pi h}[F(b,\gamma)]$$

Recall that $F(b,\gamma) < 0$ making $\Delta\sigma$ a negative magnetoconductance expected for WAL. We can thus justify simply using the WS model Eq. (2) with an opposite sign to describe WAL in our topological system:

$$\Delta\sigma(B) = \alpha \frac{e^2}{\pi h(1+\gamma^2)} |F(b,\gamma)| + \beta B^2 \quad Eq.(4)$$

$\alpha = 1/2$ per WAL channel. Previous work on III-V quantum wells have discussed a simplified implementation of the WS model[36] that we use to fit our data. Only two fit parameters are needed, $\alpha$ and $\gamma$. $\alpha$ has the same meaning as for HLN. $\gamma$ yields the coherence time $\tau_\phi$. The next section of the paper shows the results and analysis obtained for two samples QW2 and QW4.

### D. Comparison between the two models

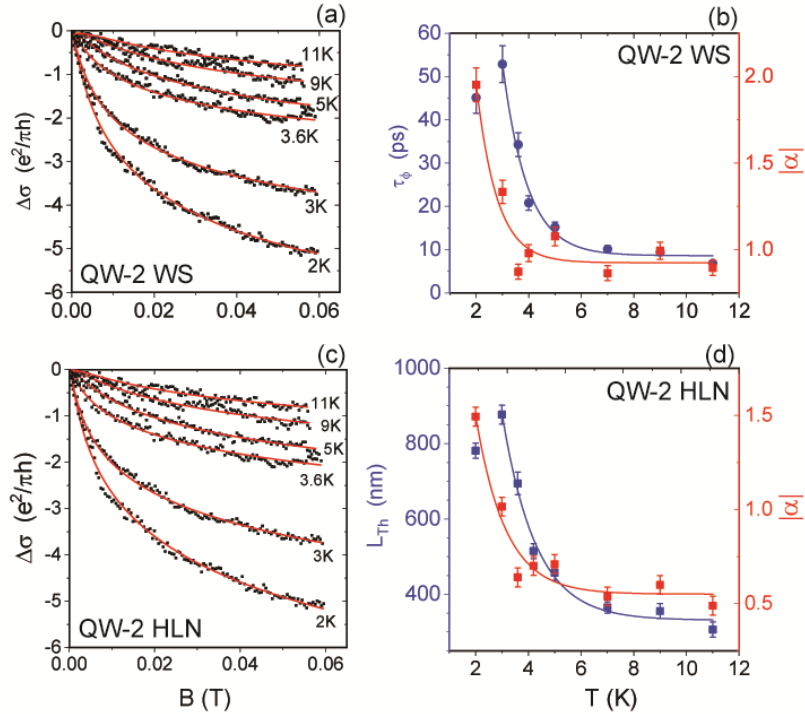

**FIG 3.** WAL analysis for QW-2. (a) WAL data at small field <0.06T ($\approx 0.5 B_{tr}$) from 2K to 10K. The solid red lines are the fitting curves obtained by the WS model and the points represent experimental data. (b) Fit parameters obtained from WS fitting. (c) Same as (a) with the fitting curve obtained by the HLN model. (d) Fit parameters obtained from HLN. $\beta$ is fixed at 0 for both WS and HLN. The lines in (b) and (d) are a guide for the eye.

Fig. 3(a) shows the curve fit by WS (Eq. (4)) performed on the WAL cusp observed between 2K and 11K in QW-2. The fit parameters from the WS model Eq. (4) are shown in Fig. 3(b). $\beta$ is neglected in this case, since the parabolic variation of the MR is extremely small below 0.06T as seen in Fig. 2(b). The coherence time increases with decreasing temperature up to 3K, indicating a decreasing inelastic scattering rate. A possible saturation is observed between 2K and 3K. A maximum coherence time close to 53ps is attained at 3K. Fig. 3(c) shows the same data analyzed using the HLN model Eq. (2), with $\beta$ also neglected. For QW-2, $B_{tr} = 1220\ Oe$, the applied field is only half of that value, limiting the validity of the HLN model (Eq. (1)). The fit parameters are shown in Fig. 3(d). $L_{Th}$ reaches a maximum close to 880nm and then saturates. The behavior of $L_{Th}$ mirrors the decay of $\tau_\phi$ as inelastic scattering is – as expected – enhanced with increasing temperature.

Fig. 4 shows an identical analysis performed for QW-4 between 2K and 10K. Fig. 4(a) shows curve fits obtained using the WS model Eq. (4), up to 600Oe, with the resulting fit parameters shown in Fig. 4(b). $\beta$ is fixed to $(820\pm30)\ e^2/\pi h(\text{Tesla})^2$ at all temperatures (see Fig. 2(d)). The decay of $\tau_\phi$ is again evident. A maximum that exceeds 100ps is obtained at 3K followed by a possible saturation. A curve fit using the HLN model is shown in Fig. 4(c), and the fit parameters are plotted in fig. 4(d). A similar enhancement of the coherence length is observed up to a possible maximum at 3K.

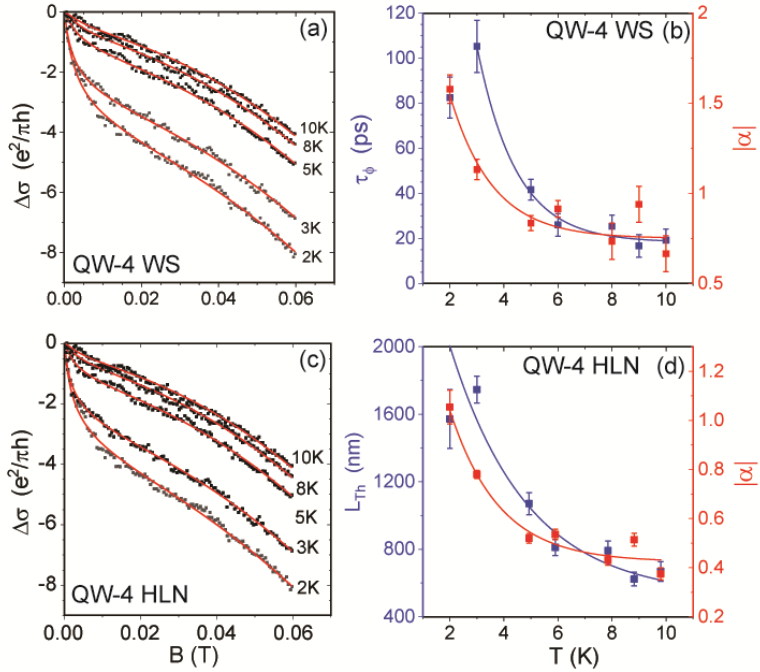

**FIG 4.** WAL analysis for QW-4. (a) WAL data at small field <0.06T ($\approx 2.4 B_{tr}$) from 2K to 10K. The solid red lines are the fitting curves obtained by the WS model and the points represent experimental data. (b) Fit parameters obtained from WS fitting. (c) Same data as (a) with the fitting curve obtained by the HLN model. (d) Fit parameters obtained from HLN. β is fixed to the value obtained in Fig. 2 for both WS and HLN. The lines in (b) and (d) are a guide for the eye.

The disagreement between the two values of $\alpha$ recovered from the two models (Fig. 3(b) compared to 3(d) and Fig. 4(b) compared to 4(d)) is expected as discussed by Wittmann and Schmid in their work when the condition b<<1 is not satisfied.[23] We have established that both models are expected to yield $|\alpha| = \frac{1}{2}$ per independent WAL contribution. It thus evident from this, that HLN underestimates α. This fact cannot be overlooked for TCI-QWs, where crystalline symmetry inherently yields multiple WAL contributions resulting from valley degeneracy. Particularly, in future studies on quantum coherent transport in high mobility TCI QWs, the HLN model may be inadequate to properly describe the valley degeneracy and the QW subbands.

In Fig. 5(a,b), the coherence length $L_{Th} = \sqrt{D\tau_\phi}$ extracted from WS is plotted in comparison with that obtained from HLN model. The Thouless length extract from WS is at least 30% lower than what is found using HLN. The result from HLN is consistently outside the error estimated for the WS fit parameters. In both QWs, the lack of agreement likely results from $B$ being comparable to $B_{tr}$.

### E. Comparison between the two samples

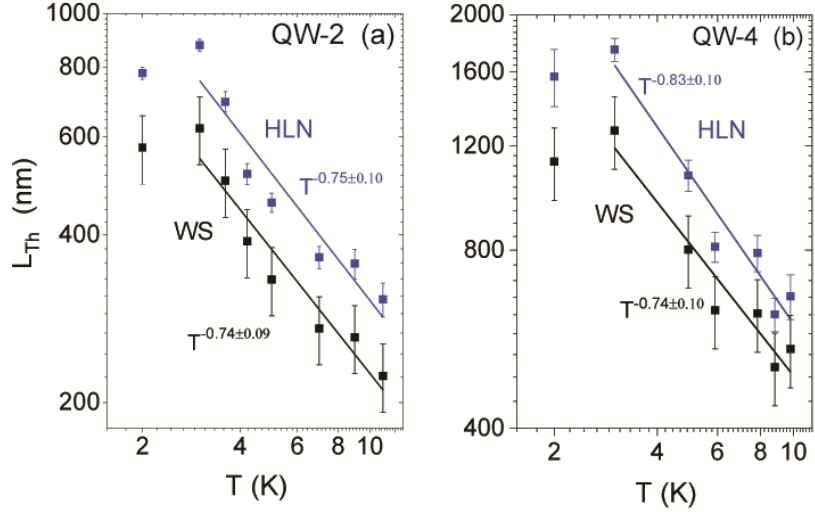

**FIG 5.** Coherence length L$_{Th}$ as a function of temperature obtained from each model for QW-2 (a) and QW-4 (b) by fitting up to 0.06T. The uncertainty on L$_{Th}$ in the WS model is dominated by the uncertainty on the diffusion constant discussed in appendix B.

When comparing the two samples, it is evident that QW4 achieves a large Thouless length exceeding 1μm. Thus, the improved growth of topological QWs allows to reduce impurities that cause both elastic and inelastic scattering. Regardless of the model used, a power law decay is obtained such that $L_{th} \sim T^{-p}$ with $p \approx 0.7 \sim 0.8$. We conclude that with the improvement of sample quality we approach the limit of validity of models that assume fully diffusive transport ($L_e \ll L_{th}$ or $B <, B_{tr}$) in topological systems, such as the HLN model. In this limit, the extracted value of $L_{th}$ is not reliable, however its decay versus temperature still agrees with WS model. A decay exponent $p \approx 0.7 \sim 0.8$ in a QW suggests the coexistence of multiple dephasing mechanisms. In most 2D systems, $p = 0.5$ is found. It is explained by the work of Altshuler et al. that links decoherence at low temperature to Nyquist noise from the carrier bath impacting the single particle:[38,39]

$$\tau_\phi^{-1} = \frac{D}{L_{Th}^2} \sim \frac{k_B T}{2\pi N D \hbar^2} \ln(\pi N D \hbar)$$

$N$ is density-of-states at the Fermi level. Electron-phonon scattering[40,41,34] and non-momentum conserving electron-electron scattering in the clean limit,[42,43,34] are both known to yield $L_{Th} \sim T^{-1}$. Their coexistence with the $T^{-1/2}$ exponent rule can yield an intermediate decay exponent. Decay exponents exceeding $p =$

0.5 have been reported in previous measurements on Pb$_{1-x}$Sn$_x$Se with lower mobility and high carrier density,[44,15] ruling out a mechanism that is highly density dependent and favoring electron-phonon scattering as a possible explanation. A saturation of $L_{Th}$ is also observed at 2K, however, more measurements at low temperature are needed to further understand this mechanism.

In both QWs, $\alpha$ is seen to be strongly enhanced below 4K, despite the saturation of $L_{Th}$, suggesting that multiple transport channels start contributing WAL. It is worth noting that in our TCI QWs, in addition to valley degeneracy, multiple subbands can be occupied even at low density as seen in the band structure calculated in Appendix B and shown in Fig. 6. If the inter-subband (or intervalley) scattering time is longer than the coherence time, an independent contribution to WAL from each subband (valley) is observed and $|\alpha|$ gets larger. Consequently, one can conclude that one of these two scattering times is also getting enhanced at low temperature, and starts to exceed $\tau_\phi$ between 3K and 4K. $|\alpha|$ is larger in QW-2, regardless of model, while at the same time, more subbands are partially occupied in QW2, the sample with the higher carrier density (Fig (6)). We thus hypothesize that the large $|\alpha|$ and its increase are more likely due to the presence of multiple subbands all yielding WAL, however the proper understanding of how trivial QW subbands in Pb$_{1-x}$Sn$_x$Se yield WAL still requires further experiments.[15]

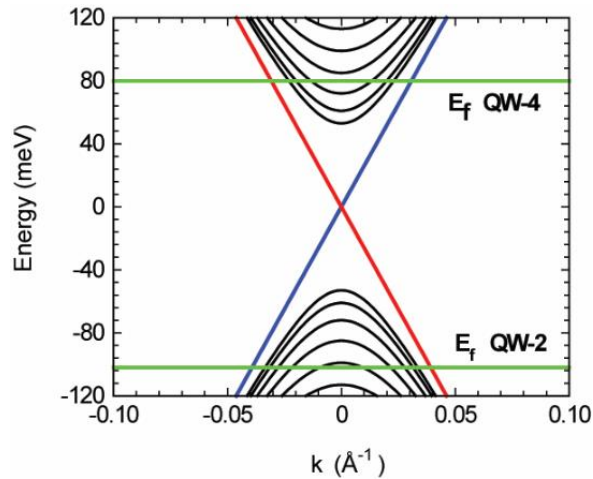

**FIG 6.** Band structure computed using the k.p formalism detailed in Appendix B and the band alignment shown in Fig. 1(a). QW subbands are shown in black and the topological interface states are shown in red and blue. The Fermi level position of QW-2 and QW-4 is shown in green. The bands are offset compared to the model parameters to have the midgap lie at zero energy. These energy bands are 4-fold valley degenerate in TCIs.

### F. Comparison with previous work

Lastly, table 2 shows a systematic comparison between our results and transport parameters extracted from previous work on TCIs and trivial IV-VI systems. We highlight that the mobility is at least an order of magnitude higher than in any previously reported TCI single QW and the carrier density is a factor of 2 to 4 lower than in Pb$_{1-x}$Sn$_x$Se on STO (see table 2). A dramatic improvement of the transport properties of TCI single QWs is thus achieved approaching what has been realized in multiquantum wells and bulk materials.[16,45] While elastic scattering is dramatically reduced compared to other works, the inelastic scattering length is comparable to what was recently measured by Kazakov et al.[15] This corroborates the reasoning that phonons or morphological defects (domain boundaries, surface scattering…) play a role in limiting quantum coherence in Pb$_{1-x}$Sn$_x$Se.

| Reference | Density [cm$^{-2}$] | Mobility [cm$^2$/Vs] | Thouless length [nm] |
|---|---|---|---|
| This work – QW-4 | **1.3 × 10$^{13}$** | **8600** | 1250 (3K) – WS<br>1750 (3K) – HLN |
| This work – QW-3 | **6.6 × 10$^{12}$** | **11300** | |
| [8] SnTe (40nm) on BaF$_2$ | >4×10$^{13}$ | <800 | 500 (2K) |
| [44] Pb$_{1-x}$Sn$_x$Se on STO | 2×10$^{13}$ | <100 | 350 (2K) |
| [9] SnTe (ultrathin) on STO | 5×10$^{14}$ | <100 | 200 (1.8K) |
| [10] SnTe on BaF$_2$ | 5×10$^{12}$, 1.7×10$^{14}$ | 400, 900 | 200, 635 (4K) |
| [15] Pb$_{0.76}$Sn$_{0.24}$Se (50nm) | 8.5×10$^{13}$ | 930 | 1830 (1.5K) - HLN |

**Table 2.** Carrier density, mobility and Thouless length obtained from different works on TCI QWs,

### III. Conclusion

In summary, we have grown high-quality Pb$_{0.7}$Sn$_{0.3}$Se/Pb$_{0.9}$Eu$_{0.1}$Se single quantum wells and reached a regime where the HLN model is no longer valid. We implement the WS model – modified to apply to the limit of very strong spin-orbit coupling - to extract the inelastic scattering parameters, which gives us access to the inelastic scattering time, in addition to the Thouless length. Although previously considered for HgTe,[27,46] models that are valid beyond the diffusion limit have not yet been employed in 3D topological insulators. Our work shows that in high mobility TIs, the HLN model can underestimate valley and subband degeneracy, thereby justifying our use of the WS model. More experiments are needed to further develop an understanding of the impact of valley and subband degeneracy on WAL in TCIs using the model developed here. Owing to our improved mobility, we have also reach a coherence time that exceeds 100ps at 3K. However, it appears that that electron-phonon scattering, or morphological defects play a key role in limiting it.

Despite this limitation, our work has achieved needed progress making single TCI QWs with high mobility available for future studies at high magnetic field. Even more importantly, we achieve this in a structure that hosts a type-I topological band alignment (Fig. 1(a)), in contrast with the broken gap alignment of the SnTe-PbTe interface.[47] With this ideal band alignment and the improved transport characteristics realized in a single well, our work enables future transport measurements in the Hall quantized regime of TCIs[54] and in the second order topological insulating regime predicted in strained TCIs.[48]

**Appendix A. Transmission electron microscopy**

Transmission electron microscopy measurements are performed on QW-4. Energy dispersive X-ray (EDX) maps (Fig. 7(a-c)) show the distribution of Sn, Eu and Pb across the heterostructure. It is evident that Sn in concentrated in well. The EDX measurements allows us to estimate the position of the barrier/well interface. A zoom-in (Fig. 7(d)) at an interface confirms that the two barriers and the well are indeed pseudomorphic.

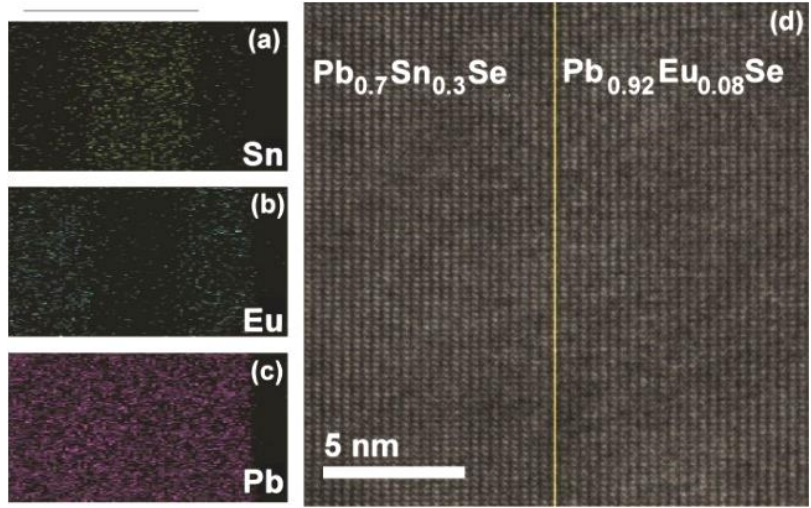

**FIG 7.** (a-c) EDX maps of QW-4 in the vicinity of the well. The scale bar is 100nm. (d) Real space TEM image of QW top interface.

**Appendix B. Band structure k.p envelope function model**

We use a k.p envelope function model to compute the band structure of our TCI QW.[49,50,16] The band structure of $Pb_{1-x}Sn_xSe$ in the bulk is comprised of the conduction band $L_6^-$ and the valence band $L_6^+$ which have opposite parity.[51] The heterostructure is comprised of a QW of $Pb_{0.7}Sn_{0.3}Se$ sandwiched between two layers of $Pb_{0.9}Eu_{0.08}Se$ grown along the [111] direction (referred to as the $\hat{z}$ direction) as shown in Fig. 1(a). The Sn/Pb ratio is fixd at x=0.3 as determined from X-ray diffraction measurements on the QW and control bulk samples. The well hosts an inverted band structure while the barrier hosts non-inverted levels. [52,53,17,26]

In this structure, a quasi-symmetric potential barrier of magnitude $V$ confines electrons in the well.[16] Since the band gap ($-\varepsilon_A$) of the well material is negative, an inverted junction is created at the interfaces and topological Dirac states emerge at this interface. The 4-band k.p envelop function model described here is used to compute the band dispersion of the topological and trivial states of the QW.[16] The thickness of the well along the growth axis is $d$ and the point $z = 0$ is chosen in the middle of the well so that the confining potential is an even function. In the setup illustrated in Fig. 1(a), zero energy is taken to be at the bottom of the bulk conduction band. If we consider only the nearest bands and spin and neglect far bands the Hamiltonian can be expressed in the basis $|L_6^+, \uparrow\rangle, |L_6^+, \downarrow\rangle, |L_6^-, \uparrow\rangle, |L_6^-, \downarrow\rangle$ [51] as:

$$\hat{H} = \begin{pmatrix} V_- & 0 & \hbar v_z k_z & \hbar v_\perp(k_x - ik_y) \\ 0 & V_- & \hbar v_\perp(k_x + ik_y) & -\hbar v_z k_z \\ \hbar v_z k_z & \hbar v_\perp(k_x - ik_y) & -\varepsilon_A + V_+ & 0 \\ \hbar v_\perp(k_x + ik_y) & -\hbar v_z k_z & 0 & -\varepsilon_A + V_+ \end{pmatrix}.$$

The diagonal terms contain the step functions defined as $V_\pm(z) = \pm V$ for z outside the well, else $V_\pm(z) = 0$. $\hat{H}$ thus depends on the energy parameters $V$ and $\varepsilon_A$, the wavevector $\mathbf{k} = (k_x, k_y, k_z)$, and the empirically determined Dirac velocity $v_D$. Note that $v_z$ is the out-of-plane velocity and $v_\perp$ is the in-plane velocity, assumed to be equal, and that these values are closely related to the Kane matrix element $P$ according to $v_z = P/m_0$.

The potential barrier in the $\hat{z}$ results in confinement such that $k_z$ is not a good quantum number. Conversely, the particle is free in the $\hat{x}$ and $\hat{y}$ direction so the other two components of the wavevector are good quantum numbers. We proceed by fixing $k_x = k_y = 0$. The simplified Hamiltonian is then

$$\hat{H} = \begin{pmatrix} V_- & 0 & -i\hbar v_z \partial_z & 0 \\ 0 & V_- & 0 & i\hbar v_z \partial_z \\ -i\hbar v_z \partial_z & 0 & -\varepsilon_A + V_+ & 0 \\ 0 & i\hbar v_z \partial_z & 0 & -\varepsilon_A + V_+ \end{pmatrix}.$$

This matrix is independent of $x$ and $y$, so that in the eigenvalue equation we cancel out all but the $z$ dependent parts of the $L_6^+$ and $L_6^-$ envelope functions, $\chi_1(z)$ and $\chi_2(z)$. The potential is even in $z$, ensuring that the $\chi(z)$ functions in the well are alternatively even and odd. Outside the well, we require they take the evanescent form $\chi(z) = Be^{-\rho(|z|-d/2)}$ so that the wavefunction will be normalizable. Furthermore, since all the nonzero matrix elements are purely $\langle \uparrow |\hat{H}| \uparrow \rangle$ or $\langle \downarrow |\hat{H}| \downarrow \rangle$, we may separate the problem by spin into

$$\hat{H}_\downarrow \vec{\Psi}_n = E\vec{\Psi}_n \text{ for } \hat{H}_\downarrow = \begin{pmatrix} V_- & -i\hbar v_z \partial_z \\ -i\hbar v_z \partial_z & -\varepsilon_A + V_+ \end{pmatrix} \text{ and } \vec{\Psi}_n = \begin{pmatrix} \chi_1^n \\ \chi_2^n \end{pmatrix}$$

$$\text{and } \hat{H}_\uparrow \vec{\varphi}_n = E\vec{\varphi}_n \text{ for } \hat{H}_\uparrow = \begin{pmatrix} V_- & i\hbar v_z \partial_z \\ i\hbar v_z \partial_z & -\varepsilon_A + V_+ \end{pmatrix} \text{ and } \vec{\varphi}_n = \begin{pmatrix} \chi_1^n \\ -\chi_2^n \end{pmatrix}$$

Here the index $n$ refers to the energy level. Enforcement of the appropriate boundary conditions at $z = \frac{d}{2}$ for the even case $\chi_1 = A\cos(k_z z)$ gives two equations,

$$A\cos\left(k_z \frac{d}{2}\right) = B \quad \text{and} \quad \frac{-Ak_z}{-E-\varepsilon_A}\sin\left(\frac{k_z d}{2}\right) = \frac{-B\rho}{-E-\varepsilon_A+V}$$

which can be divided to obtain a single equation which can be solved for $E$,

$$\tan\left(\frac{k_z d}{2}\right) = \frac{\rho(E+\varepsilon_A)}{k_z(E+\varepsilon_A-V)} \quad \text{for } \chi_1 \text{ even.} \quad \text{Eq. (1)}$$

This process is easily repeated to obtain three additional equations,

$$\cot\left(\frac{k_z d}{2}\right) = -\frac{\rho(E+\varepsilon_A)}{k_z(E+\varepsilon_A-V)} \quad \text{for } \chi_1 \text{ odd,} \quad \text{Eq. (2)}$$

$$\tanh\left(\frac{-ik_z d}{2}\right) = \frac{\rho(E+\varepsilon_A)}{ik_z(E+\varepsilon_A-V)} \quad \text{for } \chi_1 \text{ even and } k_z \text{ imaginary,} \quad \text{Eq. (3)}$$

$$\text{and } \coth\left(\frac{-ik_z d}{2}\right) = \frac{\rho(E+\varepsilon_A)}{ik_z(E+\varepsilon_A-V)} \quad \text{for } \chi_1 \text{ odd and } k_z \text{ imaginary.} \quad \text{Eq. (4)}$$

The appropriate wavenumbers are given by $k_z = \frac{1}{\hbar v_z}\sqrt{E(E+\varepsilon_A)}$ inside the well and $\rho = \frac{1}{\hbar v_z}\sqrt{(E+V)(-E-\varepsilon_A+V)}$ inside the barrier.

Note that Eq. 3 and 4 are valid in the energy regime $-\varepsilon_A < E < 0$ and each can admit only one solution representing the topological interface states of the system. Eq. 1 and 2 are valid at any other energy and represents the confined levels of the QW. This approach is identical to the one utilized in ref.[16] to determine the energy levels of Pb$_{1-x}$Sn$_x$Se QW.

The energy dispersion of the QW states is needed to determine the diffusion constant. A systematic perturbation scheme is usually applied to the QW states to determine the dispersion.[16] The result is evidently a set of massive Dirac bands and one massless Dirac state representing the topological state. In ref [16], one can notice that the Dirac velocity that determines the dispersion of the QW states is almost equal to that of the bulk bands.[16] Therefore, we can justify, a posteriori, simply using

$$E_i(k) = \sqrt{\Delta_i^2 + (\hbar v_D k)^2}$$

With $v_D \approx 3.95 \times 10^5 m/s$ to compute the energy dispersion of all the QW levels including the topological interface states. $\Delta_i$ is the energy separation between each subband edge and the midgap. It is equal to 0 for the topological states. Using the following input parameters for the model, $d = 50nm$, $\mathcal{E}_A = -100meV$, $v_D = v_z = 3.95 \times 10^5 m/s$ [26] and $V = V^{+/-} = \pm 250meV$ [54], we obtain the band dispersion shown in Fig. 6.

**Appendix C. The diffusion constant in a multiband system**

In order to determine the diffusion constant in a multiband system, a good knowledge of the band structure is required. This is a consequence of the fact that

$$D = \frac{1}{2} v_f^2 \tau$$

There $v_f$ is the Fermi velocity. From our measured carrier density, we are able to determine the Fermi energy with fair reliability. The scattering time $\tau$ can be obtained directly from the mobility. Below, we show a detailed description of how $D$ is determined.

Given the band structure calculated above, we can calculate Fermi energy $E_f$ based on measured carrier density $n$ by

$$E_i(k) = \sqrt{\Delta_i^2 + (\hbar v_D k)^2}$$

$$n_i = \frac{k^2}{2\pi} \quad with \quad n_{experimental} = 4 \sum n_i$$

The factor of 4 is due to the 4-fold valley degeneracy of TCIs. $i$ is the subband index. The effective mass $m$ is energy dependent in non-parabolic and Dirac semiconductors. $m$ at the Fermi energy can be calculated from:

$$\frac{1}{m} = \frac{2}{\hbar^2} \frac{dE}{d(k^2)} = \frac{v_D^2}{E_f}$$

We can then calculate effective Fermi velocity at $E_f$ for each subband simply using:

$$v_i(k) = \frac{\hbar k}{m}$$

The effective Fermi velocity is then a weighted average velocity over all partially occupied subbands:

$$v_f = \sum \frac{v_i \cdot N_i}{N_{tot}}$$

Here $N_i$ is the density of states and $v_i$ the Fermi velocity at the Fermi level for each partially occupied band, and $N_{tot}$ is the sum over all density of states, as discussed in previous experimental work on HgTe QW[27] and in general for any multiband system.[28] IV-VI materials host a nearly-ideal massive Dirac dispersion. $v_f$ is then a simple average over all partially occupied bands.

The elastic scattering time can now be extracted from the experimental measurement of the mobility:

$$\tau_e = \frac{m\mu}{e}$$

We can now determine $L_e = v_f \tau_e$ and the diffusion constant $D$. We use this computed $L_e$ to determine $B_{tr}$ as shown in table 1. For QW-4, we find the effective mass $m \approx 0.089 m_0$ and $v_f \approx (2.7 \pm 0.2) \times 10^5 m/s$. From the mobility, we get $\tau_e \approx 0.42 ps$, so elastic scattering length $L_e = v_f \tau_e \approx (114 \pm 8) nm$ and $D \approx (160 \pm 24) cm^2/s$. For QW-2, $m \approx 0.13 m_0$ and $v_f \approx (2.8 \pm 0.2) \times 10^5 m/s$. The Hall mobility thus gives, $\tau_e \approx 0.18 ps$, $L_e = (52 \pm 4) nm$ and $D = (73 \pm 10) cm^2/s$.

The uncertainty on $v_f$ is dominantly due to the uncertainty on the composition of the well, i.e. the bulk band gap on Pb$_{1-x}$Sn$_x$Se layer (with x=0.3±0.03).

### Appendix D. The strong-spin orbit coupling limit of quantum interference models

We further justify our choice to employ the Wittman and Schmid model with an inverted sign beyond the fully diffusive limit as follows

For the WS model,

$$\Delta \sigma_{WL}(b) = -\frac{2e^2}{\pi \hbar} D \int dt_0 \, W_{t_o} e^{-t_0/\tau_\phi}$$

$W_{t_o}$ is related to the probability of return of a charge carrier to its starting point at time 0 in a given time $t_0$. By the same reasoning considered by Zduniak et al.[37] Germanenko et al. [55,56] and Abrikosov[57] in the presence of spin-orbit relaxation,

$$\Delta \sigma_{WAL}(b) = -\frac{2e^2}{\pi \hbar} D \int dt_0 \, W_{t_o} \left[ e^{-t_0/\tau_{SO}} + \frac{1}{2} e^{-2t_0/\tau_{SO}} - \frac{1}{2} \right] e^{-t_0/\tau_\phi}$$

$e^{-t_0/\tau_{SO}}$ and $e^{-2t_0/\tau_{SO}}$ represent spin relaxation in the triplet channel for the in-plane and out of plane directions.

$$\Delta \sigma_{WAL}(b) = -\frac{2e^2}{\pi \hbar} D \int W_{t_o} \left[ e^{-t_0\left(\frac{1}{\tau_\phi}+\frac{1}{\tau_{SO}}\right)} + \frac{1}{2} e^{-t_0\left(\frac{1}{\tau_\phi}+\frac{2}{\tau_{SO}}\right)} - \frac{1}{2} e^{-t_0/\tau_\phi} \right] dt_0$$

For a topological system $\tau_{SO}$ is very small $(\tau_{SO} \sim \tau_e) \ll \tau_\phi$

$$\text{So, } e^{-t_0\left(\frac{1}{\tau_\phi}+\frac{1}{\tau_{SO}}\right)} \sim e^{-t_0\left(\frac{1}{\tau_\phi}+\frac{2}{\tau_{SO}}\right)} \ll e^{-t_0/\tau_\phi}$$

$$\Delta\sigma_{WAL}(b) = \frac{2e^2}{\pi\hbar}D\int W_{t_o}\left[\frac{1}{2}e^{-t_0/\tau_\phi}\right]dt = -\frac{1}{2}\Delta\sigma_{WL}$$

The multiplicative factor of -1/2 in the limit of strong spin-orbit coupling appears regardless of model used for WAL. It is worthwhile noting, that we simply treat topological surface states as states with very strong spin-orbit coupling, strong enough to neglect the spin-triplet terms considered by Zduniak et al. For near-ideal topological surface states with spin locked to the momentum indeed the triplet channel can be completely neglected as shown in ref [21,18].

Lastly we would like to note that apart from the WS model, Kawabata has provided a theory that is also valid beyond the diffusion limit, however, it was argued in subsequent literature that Kawabata's expression included additional terms from N=1 and N=2 loops in Eq. (1) that do not yield any quantum interference.[58] For this reason, the Kawabata model was argued to be incomplete. Dmitriev et al. have also treated the problem beyond the diffusion limit and included non-backscattering contributions, however, their expression is more cumbersome to implement and did not included spin relaxation.[59] The treatment of WS followed by the work of ref [37,36] makes the WS model ideal to consider in our context due to its relative simplicity and its assumptions.

### **Acknowledgements**

The work is supported by NSF-DMR 1905277 and partly by a seed grant from Notre Dame Nanoscience and Technology (NDnano). The Material Characterization Facility is funded by the Sustainable Energy Initiative (SEI), which is part of the Center for Sustainable Energy at Notre Dame (ND Energy). We also acknowledge support from the Notre Dame Integrated Imaging Facility.